\documentclass[a4paper,fleqn]{cas-dc}

\usepackage[numbers]{natbib}
\usepackage{ragged2e}
\usepackage{float}
\usepackage{graphicx}
\def\tsc#1{\csdef{#1}{\textsc{\lowercase{#1}}\xspace}}
\tsc{WGM}
\tsc{QE}

\begin{document}
\let\WriteBookmarks\relax
\def\floatpagepagefraction{1}
\def\textpagefraction{.001}

\shorttitle{Transparency of Deep Neural Networks for Medical Image Analysis: A Review of Interpretability Methods}    

\shortauthors{Z. Salahuddin et al.}  
\title [mode = title]{Transparency of Deep Neural Networks for Medical Image Analysis: A Review of Interpretability Methods}

\author[1]{Zohaib Salahuddin}[orcid=0000-0002-9900-329X]

\cormark[1]
\ead{z.salahuddin@maastrichtuniversity.nl}
\cormark[1]
\credit{Conceptualization of this study, Methodology, Software}
\affiliation[1]{organization={The D-Lab, Department of Precision Medicine},
            addressline={GROW – School for Oncology and Developmental Biology, Maastricht University}, 
            city={Maastricht},
            country={The Netherlands}}

\affiliation[2]{organization={Department of Radiology and Nuclear Medicine},
            addressline={GROW – School for Oncology and Developmental Biology, Maastricht University Medical Centre+}, 
            city={Maastricht},
            country={The Netherlands}}

\author[1,2]{Henry C Woodruff}
\author[1]{Avishek Chatterjee}
\author[1,2]{Philippe Lambin}

\cortext[cor1]{Corresponding author}

\begin{abstract}
Artificial Intelligence (AI) has emerged as a useful aid in numerous clinical applications for diagnosis and treatment decisions. Deep neural networks have shown same or better performance than clinicians in many tasks owing to the rapid increase in the available data and computational power. In order to conform to the principles of trustworthy AI, it is essential that the AI system be transparent, robust, fair and ensure accountability. Current deep neural solutions are referred to as black-boxes due to a lack of understanding of the specifics concerning the decision making process. Therefore, there is a need to ensure interpretability of deep neural networks before they can be incorporated in the routine clinical workflow. In this narrative review, we utilized systematic keyword searches and domain expertise to identify nine different types of interpretability methods that have been used for understanding deep learning models for medical image analysis applications based on the type of generated explanations and technical similarities. Furthermore, we report the progress made towards evaluating the explanations produced by various interpretability methods. Finally we discuss limitations, provide guidelines for using interpretability methods and future directions concerning the interpretability of deep neural networks for medical imaging analysis.
\end{abstract}


\maketitle
\credit{Conceptualization of this study, Methodology}

\section{Introduction}
Medical imaging plays a key role in modern medicine as it allows for the non-invasive visualization of internal structures and metabolic processes of the human body in detail. This aids in disease diagnostics, treatment planning and treatment follow-up by adding potentially informative data in the form of patient-specific disease characteristics \cite{Lambin2013PredictingOI, Aerts2014}. The amount of healthcare imaging data is rapidly increasing due to advances in hardware, the increase in population, decrease in cost, and the awareness of the utility of the imaging modalities \cite{Smith-Bindman2008}. This adds to the increasing difficulty for radiologists and clinicians to cope with the mounting burden of analyzing the large amounts of available data from disparate data sources, and studies have highlighted sometimes considerable inter-observer variability when performing various clinical imaging tasks \cite{Saha2016InterobserverVI}. It follows that there is an evolving need for tools that can aid in diagnosis and decision making.

\begin{figure*}[h!]
\centerline{\includegraphics[width=\textwidth]{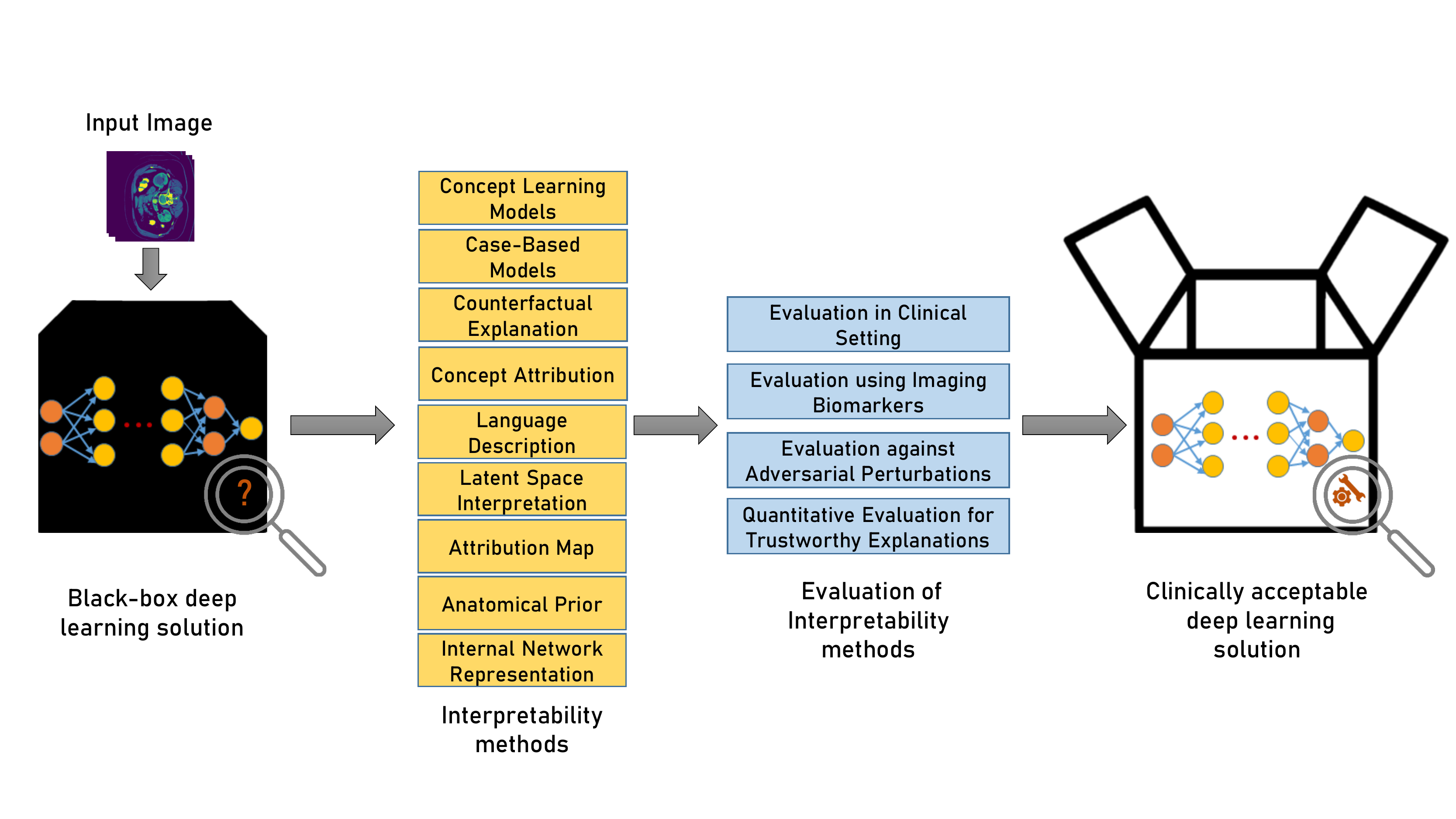}}
\caption{Overview of Interpretability methods for DL solutions in medical image analysis. The black-box DL solution can be made more desirable for clinical use by incorporating interpretability during the design or execution phase. The evaluation of explanations is important to ensure robustness of the interpretability methods.}
\label{figure: overview}
\end{figure*}

Recent advances in artificial intelligence (AI) have started permeating into healthcare, among those the so-called deep learning (DL) methods, which consist of non-linear modules that are able to learn multiple levels of representations automatically from  high dimensional data without any need of explicit feature engineering by humans \cite{LeCun2015,SCHMIDHUBER201585}. Deep learning methods, in particular convolutional neural networks (CNN),  first rose to prominence in 2012 when they demonstrated state-of-the-art performance on the ImageNet Large-Scale Visual Recognition Challenge (ILSVRC)  \cite{krizhevsky2012imagenet}, and have since become the \textit{de facto} standard for solving computer vision problems. Deep learning has also demonstrated state-of-the-art performance on many medical imaging challenges related to classification \cite{classificationchallenge1, BACHGrandChallenge2021}, segmentation \cite{ZHUANG2019101537, Yang2018AutosegmentationFT, m_m_segmentation2021} and other tasks \cite{CA_CNN_21}.

Despite their promising performance, the adoption of deep learning networks in healthcare has been slow. A contributing factor to this is that more insight into the inner workings of AI tools is needed before translation into the clinic is widely successful \cite{lipton2017doctor}. DL models are referred to as black-boxes due to their opaque nature as the interpretation of the inner states of the model is not as straightforward as with standard object-oriented code or decision trees \cite{Durn2021WhoIA}. Furthermore, the European Union’s General Data Protection Regulation (GDPR) law requires an explanation of the decision making process of the algorithm in order to make it transparent before it can be utilized for patient-care \cite{temme2017algorithms}. 

The primary purpose of AI tools for medical imaging is to aid clinicians in their decision  making by combining multiple factors into a model that returns an actionable output. Without any explanation of this output, the utility of the model is limited as it does not unveil the reasoning process, the limitations, and biases. Interpretability of DL systems can not only unravel any faulty processes within the algorithms, but also enables the discovery of other important information in the imaging data that otherwise might go unnoticed. Unravelling the black-box nature of the DL systems is not only a legal and ethical requirement but it is also essential for fostering clinical trust and for troubleshooting systems. Moreover, interpretability methods can also reveal new imaging biomarkers in an effort to understand the specifics of the DL model.

Explainable artificial intelligence (XAI) refers to AI solutions that can provide some details about their functioning in a way that is understandable to the end-users \cite{BARREDOARRIETA202082}. As the field is still in its infancy, there is no general consensus on the definition of the terms interpretability and explainability in this context, and recently various definitions have been proposed for these terms \cite{BARREDOARRIETA202082, Rudin2019, Lipton2018TheMO}. For the purpose of this work, we do not differentiate between interpretability and explainability as in \cite{Carvalho2019MachineLI}. We define interpretability of deep neural networks simply as any attempt to provide explanation about the decision making process of the model in a way that is understandable for the end-users. It refers to any technique that makes an attempt to answer the question "Why is the model making this prediction?" for the medical image analysis tasks.

This review is structured as follows. In section 2, we review and discuss the interpretability methods for understanding DL models trained to solve medical image analysis problems. Section 3 discusses how the explanations produced by various methods can be evaluated to ensure that they are robust, consistent, and useful for the clinicians. In section 4, we discuss the shortcomings of interpretability methods, provide guidelines for using interpretability methods, and discuss steps for the future. The objective of the review paper is to discuss the advantages, limitations, and evaluation of interpretability methods for DL models and this will ultimately help in choosing the correct interpretability methods. 

This review paper aims to answer the following research questions:
\begin{itemize}

    \item How interpretability methods can help in the incorporation of deep neural networks in the clinical workflow?
    \item What are the interpretability methods available for understanding deep neural networks for medical image analysis tasks? 
    \item How can various interpretability methods be grouped into categories?
    \item What are the advantages, limitations and challenges of different interpretability methods?
    \item What are different evaluation methods available to ensure that the explanations are trustworthy?  

\end{itemize}

The following keywords were used for literature search on Google Scholar and PubMed to find answers of the above mentioned research questions: "interpretability", "explainability", "medical imaging", "Explainable artificial intelligence", "deep learning", "evaluation", "sanity checks", "clinical validation", "imaging biomarkers", "Explainable artificial intelligence". The articles were chosen on the basis of their novelty, clinical application, and relevance to clinical workflow.

\section{Interpretability Methods}
The purpose of AI solutions is to aid clinicians in performing their work more efficiently and accurately, and not to replace them. This partnership requires trust on the side of the clinical experts, and one path to trust is understanding \cite{lipton2017doctor}. We can incorporate interpretability during the design process of the deep neural network. Post-hoc interpretability methods provide explanations for the predictions after the DL model has been trained. Interpretability methods can offer local or global explanations. Local explanations identify the attributes and features of a particular image that the DL model considers important for prediction. On the other hand, global explanations aim at identifying the common characteristics that the DL model considers when associating images with that particular class. After the literature review using systematic keyword searches, we have identified nine different types of interpretability methods for medical image analysis tasks based on the type of generated explanations and the technical similarities. Table \ref{methodssummarry} lists the different interpretability methods, their limitations, and the application of these methods for interpreting DL models.

\subsection{Concept Learning Models}
State-of-the-art DL models are trained to infer the label $y$ directly from the input image $x$. It is usually not possible for radiologists to understand the reason behind the prediction of the DL models using the same high level concepts $c$ used to arrive at a diagnosis. Hence, it is advantageous to explain the outputs of the model in terms of human interpretable concepts. This problem can be solved by first predicting these high level concepts (such as semantic features) from the input image and then using these concepts to predict the label \cite{pmlr-v119-koh20a}. These models require concepts generated by experts as an input during training-time along with the image and label. \citet{pmlr-v119-koh20a} introduced Concept Bottleneck Models for osteoarthritis grading and used 10 clinical concepts such as joint space narrowing, bone spurs, calcification etc. The clinicians can intervene at test time to change the predicted value of the clinical concept to observe the effect on the final prediction. This intervention from the clinicians resulted in an improvement in performance. This added interpretability comes at a cost of annotation of these clinical concepts, which requires time. \citet{Margeloiu2021} utilized post-hoc attribution methods to show that the concepts learned in concept bottleneck models do not relate with the input image in a meaningful way. \citet{mahinpei2021promises} showed that concept bottleneck models can be misleading as the learned encoding contains information in addition to the concept representation.

Capsule networks encode information (e.g. pose, scale, etc.)  about each feature using vectorized representation in contrast to scalar features maps used in CNNs \cite{NIPS2017_2cad8fa4}. \citet{xcap} proposed a multi-task capsule architecture called X-Caps for encoding high level visual attributes within the capsule vectors to perform explainable diagnosis for lung nodule malignancy prediction. Malignancy prediction was based entirely on concepts such as sphericity, margin, texture etc. The predicted visual attributes are regularized by a segmentation reconstruction error.  The proposed 2D explainable capsule network outperforms Hierarchical Semantic Convolutional Neural Network by \citet{SHEN201984} in terms of visual attribute modelling and malignancy score prediction. The performance of X-Caps approaches that of 3D non-explainable CNNs for malignancy prediction.  

Low level features extracted from CNN layers can be combined with predicted clinical concepts for diagnosis. Hierarchical Semantic Convolutional Neural Network (HSCNN) for lung nodule malignancy classification consists of three modules \cite{SHEN201984}. The first module extracts generalized low level features that are fed to the second module. The second module classifies the presence or absence of five nodule semantic characteristics that reflect diagnostic features relevant to radiologists (e.g. texture, margin). The third module predicts lung nodule malignancy based on the low level features from the first module and high level visual attributes from the second module. The authors showed that HSCNN achieves better results than some non-explainable 3D CNNs. 

\begin{figure}[t!]
\centerline{\includegraphics[width=1\columnwidth]{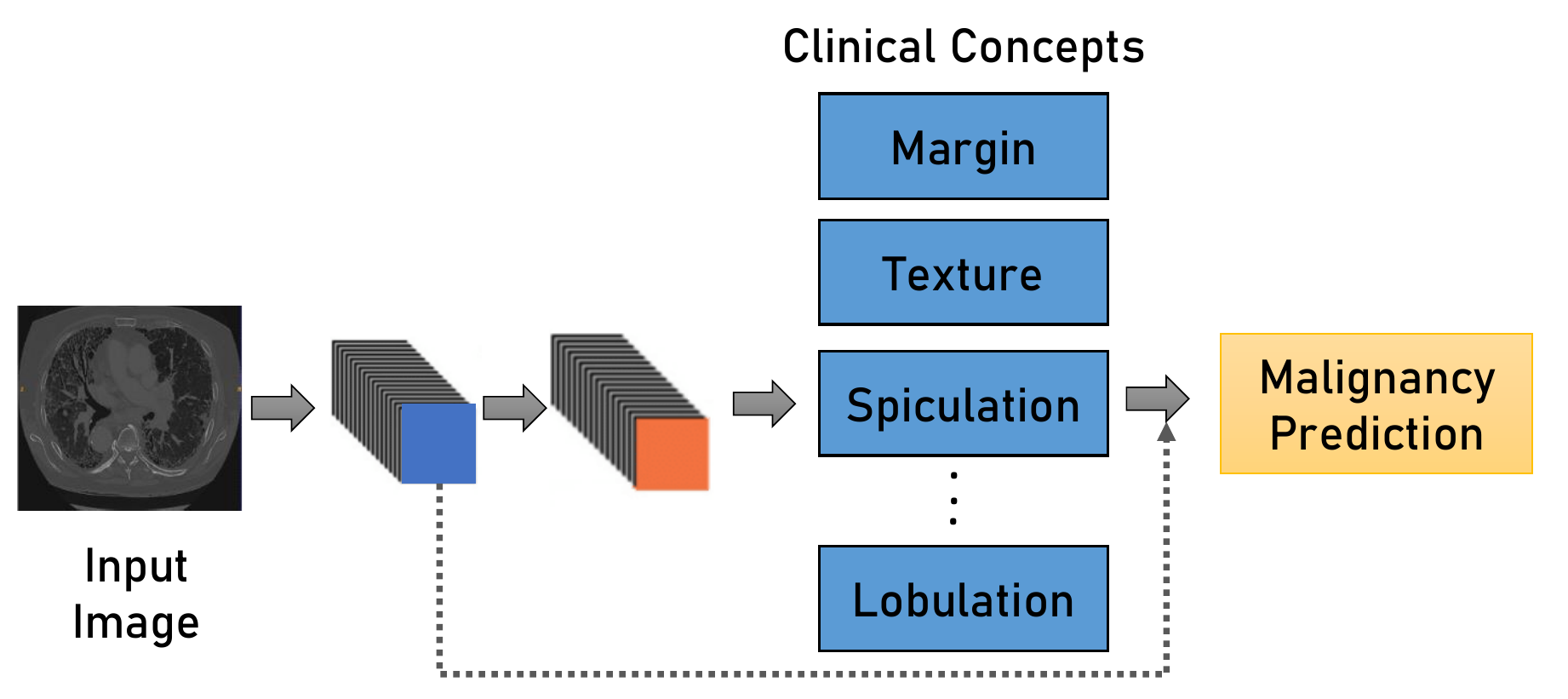}}
\caption{Concept Learning Models first predict clinical concepts and the final prediction is made using these human interpretable concepts. The final prediction is based either only on the clinical concepts or a combination of clinical concepts and other deep features (dashed line).}
\label{figure: CLM}
\end{figure}

\subsection{Case-Based Models}
\label{casebasedmodels}
Case-Based Models make predictions by comparing the features extracted from an input image against class discriminative prototypes. Prototype classification is a type of case-based reasoning that is inherently interpretable because the final predictions are made by taking a weighted sum of similarity scores between features extracted from input and prototypes \cite{LiLCR18}. \citet{ChenLTBRS19} proposed Prototypical Part Network (ProtoPNet) that consists of a convolutional layer, followed by a prototype layer and then a fully connected layer. The convolutional layer consists of a trimmed standard CNN pipeline that acts as a feature extractor. The prototype layer takes patches from the convolution layer as input and learns class discriminative prototypes during training. A similarity score is computed after comparison against each prototype. A fixed size feature map is used for comparison with prototypes. The fully connected layer then makes prediction based on these similarity scores. The convolutional layer and the prototype layer are trained first and the loss function comprises of misclassification loss, cluster cost and separation cost. In the second step, the fully connected layer is trained. The interpretability of ProtoPNet does not come at a cost of performance when compared to black-box DL models. \citet{Mohammadjafari2021Using} utilized ProtoPNet with  DenseNet121 architecture for Alzheimer's disease classification. \citet{barnett2021interpretable} used a modified version of ProtoPNet to utilize fine-grained expert annotations for mass margin classification and malignancy prediction. The prototype networks are comparatively difficult to train than regular deep neural networks because of the scarcity of available techniques for their training \cite{Rudin2019}. 

ProtoPNet uses a fixed size feature map from an input image for comparison with prototypes. Hence, there is a possibility that the patch may not be able to capture class discriminative features. XProtoNet extracts features within a dynamic area and is not constrained to a limited size for comparison with prototypes \cite{Kim_2021_CVPR}. XProtoNet achieved state-of-the-art performance for the classification of 14 diseases on public chest X-ray dataset. Generalized Prototypical Part Network (Gen-ProtoPNet) puts no constraint on the dimensions of the prototype \cite{CovidGurmail2021}. Gen-ProtoPNet was used for the classification of Normal, COVID and Pneumonia classes in chest X-ray. Negative-Positive Prototypical Part Network \linebreak (NP-ProtoPNet) is an extension of ProtoPNet that takes the association between similarity scores and incorrect class into consideration \cite{Singh2021TheseDN}. NP-ProtoPNet was utilized for the classification of COVID and Pneumonia in chest X-ray. Similarity scores in ProtoPNet can be corrupted by noise and JPEG compression artefacts \cite{hoffmann2021looks}. Similarity in latent space does not always translate to similarity in terms of human-interpretable features.

\begin{table*}[H]
\label{methodssummarry}
\caption{The table summarizes different interpretability methods, their limitations and the application of these methods for explaining DL models for medical image analysis tasks.}
\centering
\begin{tabular}{p{0.025\textwidth}p{0.15\textwidth}p{0.30\textwidth}p{0.275\textwidth}p{0.15\textwidth}}
\hline
\Centering{No.} 
&\Centering{Interpretability Methods} 
& \Centering{Description} 
& \Centering{Limitations} 
& \Centering{References}
\\ \hline

\centering{1.} 
& \Centering{Concept Learning Models}
& High level clinical concepts are first predicted and the final classification is made using these concepts.
& Additional annotation cost, learned concepts may encode information beyond the intended clinical \linebreak concepts due to information leakage.

&  \cite{pmlr-v119-koh20a}, \cite{SHEN201984}, \cite{NIPS2017_2cad8fa4}, \cite{SHEN201984} \\

\Centering{2}
&\Centering{Case-Based Models}
& Class discriminative prototypes are learned and the final classification is performed by comparing features \linebreak
extracted from input images with the prototypes. 

& Susceptibility to corruption by noise and compression artefacts, difficult to train.
& \cite{LiLCR18}, \cite{Mohammadjafari2021Using}, \cite{ChenLTBRS19}, \cite{Singh2021TheseDN}, \cite{CovidGurmail2021}, \cite{Kim_2021_CVPR}\\

\Centering{3.}
&\Centering{Counterfactual Explanation}
& Input images are perturbed in a \linebreak
realistic manner to generate the \linebreak opposite prediction.
& Possibility of unrealistic perturbations to the  input images, the resolution of the generated counterfactual images is limited.

& \cite{Cohen2021GifsplanationVL}, \cite{miccaicounterimpact}, \cite{Selvaraju2019GradCAMVE}, \cite{Simonyan2014DeepIC}, \cite{Schutte2021UsingSF}, \cite{Seah2019ChestRI}, \cite{singla2021explaining}, \cite{NEURIPSICAM}, \cite{TANG2021101839}, \cite{Baumgartner2018VisualFA}\\

\Centering{4.}
&\Centering {Concept Attribution}
& Global explanations to quantify the influence of high level image concepts/radiomics features on the model predictions.
& Difficult to annotate high level clinical concepts, radiomics features used for interpretability may not be reproducible.  

& \cite{Kim2018InterpretabilityBF}, \cite{Gamble2021}, \cite{CloughTCAV2019}, \cite{CloughTCAV2019}, \cite{DTCAV2021}, \cite{Graziani2018RegressionCV}, \cite{DeepDreams2019}, \cite{Yeche2019UBSAD} \\

\Centering{5.}
&\Centering{Language Description}
& Textual justifications are provided along with the predictions. 
& Structured diagnostic reports require more annotation efforts, \linebreak duplication of training sentences during testing.

&  \cite{textualjustification2019}, \cite{Wang2018TieNetTE} \cite{Zhang2017MDNetAS}, \cite{Zhang2019PathologistlevelIW}, \cite{symbolicsegmentation}, \cite{Chowdhury2021EmergentSL}\\

\Centering{6.}
&\Centering{Latent Space Interpretation}
& The latent space is used to uncover the salient factors of variation learned in the data with respect to the clinical knowledge. Visualization of high-dimensional \linebreak latent space in two dimensions to identify similarities and outliers.

& Loss of information when the high-dimensional feature space is \linebreak projected to two dimensions. Similarity in latent space does not always translate to similarity in terms of human-interpretable features. 
&  \cite{biointerpret2020}, \cite{Yang2019DomainAgnosticLW}, \cite{Chen2020}, \cite{Biffi2020ExplainableAS}, \cite{CloughTCAV2019}, \cite{JMLR:v9:vandermaaten08a}, \cite{unbox2021}, \cite{DINSDALE2021117689}, \cite{Faust2018VisualizingHD}, \cite{latentrepresentation2020}
\\

\Centering{7.}
&\Centering{Attribution Map}
& Post-hoc explanations are provided by highlighting the regions of the input image that the model considers \linebreak important. 
& No information is provided on how the relevant regions contribute to the prediction, multiple classes can have the same regions highlighted. 

& \cite{Boehle2019LayerWiseRP}, \cite{MLCAM2018}, \cite{Camalan2021}, \cite{gradcamattr2018}, \cite{SAYRES2019552},  \cite{KERMANY20181122}, \cite{ParkinsonLime2020}, \cite{LundbergSHAP2017}, \cite{Ghorbani2020}, \cite{WICKSTROM2020101619}, \cite{imageretreival2020} \\

\Centering{8.}
&\Centering{Anatomical Prior}
& Task specific structural information is incorporated in the design process of the network. 

& Specialized clinical knowledge may be required, anatomical prior cannot be utilized for all problems. 

& \cite{midline2019}, \cite{robustshape}, \cite{attentionSAUNet2020} \\

\Centering{9.}
&\Centering{Internal Network Representation}
& Visualization of features learned by different filters in a CNN.
& The structures and patterns that different filters learn to identify are hard to interpret in medical images.

& \cite{Bau2017NetworkDQ}, \cite{natekar2020} \\
\hline

\end{tabular}
\end{table*}

\subsection{Counterfactual Explanation}
\label{counterfactexplain}
Counterfactual explanation is an image that is produced by applying minimal perturbations to the original image in order to bring a maximum change in classifier's prediction and switch the predicted class of the original image \cite{Verma2020CounterfactualEF}.  
Counterfactual explanation not only helps in identifying the diseased area but also aids in understanding the changes that need to be made in order to switch the classifier's prediction. Counterfactual images are synthesized by using Generative Adversarial Networks (GANs) \cite{goodfellow2014generative} or by perturbing the latent space of an autoencoder. Generative Adversarial Networks (GANs) \cite{Goodfellow2014GenerativeAN} are widely used for image synthesis because of their ability to model target data distribution. GANs consist of a generator and a discriminator that work in an opposing manner. GANs can be difficult to train due to loss function instability and high sensitivity to hyper-parameters. 

 \citet{Schutte2021UsingSF} proposed a method using StyleGAN2 \cite{Karras2019ASG} to generate synthetic images by minimal modification that would lead to new predictions. A generator $G$ has an intermediate latent representation $z$. An encoder $E$ is trained to retrieve the latent representation $z$ back from the generated image. A classifier $f$ can be explained by fitting a logistic regression classifier $f^{'}(z_i) = \sigma(\alpha \cdot z_i + \beta)$ in the latent space $z$ to estimate the labels predicted by $f$. The latent representation $z=E(X)$ of an input image X is translated in the direction $\alpha$ to produce a synthetic image ($X^{'} = G(z + \lambda \cdot \alpha)$) that has a lower or higher prediction depending on the value of $\lambda$. This method was validated on knee osteoarthritis severity prediction on X-ray images and tumor detection on histology images of metastatic lymph nodes. \citet{counterfactuals_image_translation} used unsupervised image-to-image translation using CycleGAN \cite{cyclegan} to generate counterfactual explanations. Translation between the two classes was applied successively to amplify the differences. A linear SVM trained using features obtained from identified novel image regions achieved performance similar to that of CNNs. This method was demonstrated using retinal images for predicting Diabetic Macular Edema. \citet{Ghandeharioun2021DISSECTDS} proposed an approach called DISSECT that jointly trains a generator, discriminator and a concept disentangler to generate a series of example with increasing degree of concepts that effect the classifier decision. This method was validated on a synthetic dataset that reflected real world characteristics of melanoma skin lesions. \citet{Baumgartner2018VisualFA} proposed an attribution method based on Wasserstein Generative Adversarial Networks (WGAN) \cite{wgan} that requires a baseline class $y$ and class of interest $x$. WGAN estimates a function $M(x_i)$ such that when it is added to image from class $x_i$, it becomes indistinguishable from class $y$ . This attribution technique was validated on a synthetic dataset and MRIs from patients with Alzheimer's disease and mild cognitive impairment.  \citet{NEURIPSICAM} proposed a framework for simultaneous classification and generation of attribution map. This method employed two variational autoencoders (VAE) for generating class relevant and irrelevant latent space for each input image. A discriminator loss on the decoder is used to extract the class irrelevant latent space. The class irrelevant latent spaces of two images from class $x$ and $y$ is swapped and fed to a generator to produce an attribution map. The class relevant latent space is used for classification and visualization of variations between classes. This method was validated on three brain imaging datasets.

The latent space of an autoencoder can be perturbed to synthesize new images but the resolution of the generated images is limited.  \citet{Cohen2021GifsplanationVL} trained an autoencoder with an Encoder $E(x)$ and a Decoder $D(z)$ in order to produce a latent representation $z$ for an input image $x$. Perturbations in the latent space for a classifier $f$ can be produced as: $z_y \: = z + \lambda \: \frac{\partial\:f(D(z))}{\partial\:z}$ to produce a new image $x_{y}^{'} = D(z_y)$ that results in a higher prediction ($f(x_{y}^{'}) \: > f(x)$). Different images $x_{y}^{'}$ can be combined together for different values of $\lambda$ to produce a gif. This method was applied to explain chest X-ray classifiers. \citet{Seah2019ChestRI} utilized Generative Visual Rationales (GVRs) for the interpretability of CNNs for predicting congestive heart failure from chest radiographs. A GAN architecture is trained on an unlabelled dataset and it is capable of generating realistic images. An autoencoder is trained to produce latent space representation. The latent space is permuted until the disease is no longer predicted. The generator trained on the unlabelled dataset then synthesizes a healthy image from the perturbed latent space. The difference of the diseased and reconstructed healthy radiograph produces the GVR.


\begin{figure*}[h!]
\centerline{\includegraphics[width=\textwidth]{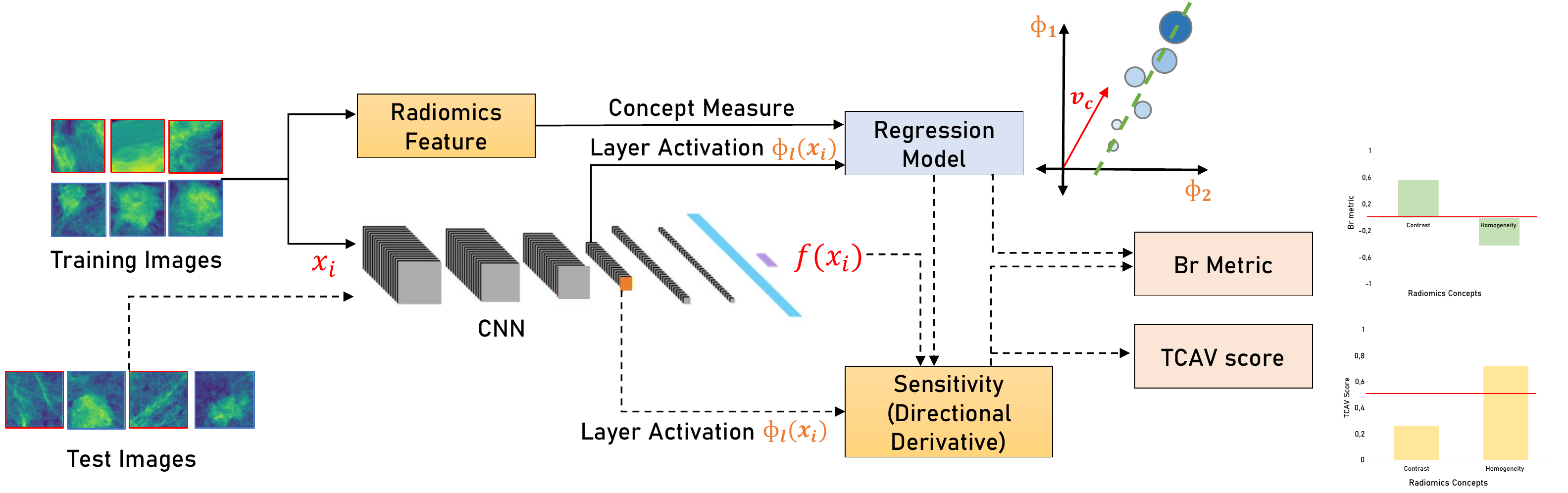}}
\caption{Regression Concept Vectors for global interpretability. A linear regression model is trained to estimate radiomics features using the activations $\phi_{l}$ of layer $l$ from a trained CNN. The influence of radiomics features on the decision of a particular class is obtained by calculating the directional derivation in the direction of increase of radiomics concepts during testing phase (dashed lines). The effect of radiomics features is quantified in terms of Bidirectional relevance (Br) metric and Testing with Concept Activation Vectors (TCAV) score. }

\label{figure: overview}
\end{figure*}

\subsection{Concept Attribution}
\label{conceptattr}
Concept attribution provides global explanations for the deep neural network in terms of high-level image concepts \cite{M2020103865}. Testing with Concept Activation Vectors (TCAVs) method quantifies the influence of a high level image feature on the decision of the model \cite{Kim2018InterpretabilityBF}. A linear classifier is trained in order to differentiate between examples containing the concept of interest and random examples. Concept Activation Vector (CAV) is orthogonal to the classification boundary of this linear classifier. TCAV method utilizes concept activation vector and directional derivative to determine the importance of a particular concept for classification in terms of TCAV score. \citet{Kim2018InterpretabilityBF} used TCAV to provide explanations for a model that predicted Diabetic Retinopathy (DR) level using a five point score. The TCAV score was high for concepts that are relevant for the diagnosis of a particular DR Level but low for concepts that are not associated with diagnosis of a particular DR level. \citet{Gamble2021} used TCAV for interpretability of DL models for breast cancer biomarker status prediction.  \citet{CloughTCAV2019} used a series of fully connected layers that process the latent space of a variational autoencoder (VAE) for cardiac disease classification in CMR images. The sensitivity of the classifier to several clinical biomarkers was evaluated using TCAV in the activation space of fully connected layer with 64 units. TCAV showed that ventricular ejection and filling rates are the most relevant features for the classifier. 

It can be challenging to create a label dataset for different concepts in order to obtain the concept activation vectors especially in the field of medical imaging. Automated Concept-based Explanation (ACE) method mitigates the need of human supervision for labelling the concepts  \cite{D-TCAV}. In ACE, segments are extracted from images of the same class at multiple resolutions. These segments are resized to the original image size. The Euclidean distance in the activation space of the bottleneck layers is used to group together similar segments as concepts. \citet{DTCAV2021} used superpixels as segments and consequently employed ACE method for automated concept discovery. These concepts were then used for TCAV to interpret different cardiac conditions from a DL model for segmentation in Cardiac Magnetic Resonance (CMR).

Radiomics converts medical images into quantitative features that can be utilized for clinical decision support \cite{Lambin2017RadiomicsTB, Lambin2012RadiomicsEM}. Radiomics features can be used as continuous-valued concepts as they are understandable by humans.  \citet{Graziani2018RegressionCV} extended TCAV so that continuous variables such as radiomics features could be used as concepts. Regression Concept Vector (RCV) in the direction of greatest increase of the concept measure is obtained by solving least square linear regression problem in the activation space of a particular layer. A Bidirectional relevance (Br) metric along with the TCAV score is used to measure the importance of the continuous-valued concept. This approach was validated on breast cancer histopathology classification. Testing with regression concept vectors identified contrast and correlation as the most relevant concepts for the classification task. \citet{Yeche2019UBSAD} used regression concept vectors to identify important radiomics features for calcification and mass classification from patches extracted from mammograms. A new metric called Unit Ball Surface Sampling metric (UBS) was introduced in-order to overcome the shortcomings of Br metric and TCAV score. UBS metric does not depend on the current layer and is indifferent to the concept's representation in feature space. 

 \citet{DeepDreams2019} proposed a method to test the robustness and sensitivity of the deep segmentation networks based on radiomic features. The network is robust to a feature if it is not sensitive to its variations. Input image is altered to produce a positive response in the direction of the steepest slope in the input space using network gradients. At each iteration, radiomic features of the segmentation mask are calculated. The network is sensitive to the radiomic features that change during this process of activation maximization. This method was applied to interpret deep neural networks for liver tumor segmentation.

\subsection{Language Description}
Deep neural networks can provide explanations in the form of text along with the predictions. The textual descriptions can consist of natural language justifications that are learned in a supervised manner or emergent language justifications that are learned in an unsupervised manner. 
\label{languagedescrip}
\subsubsection{Natural Language}
Deep Neural Networks can be trained to provide textual justifications along with the predictions \cite{Wang2018TieNetTE, Zhang2017MDNetAS}. Structured diagnostic reports along with images can be utilized for the training of such networks. It is challenging to train networks that provide natural language descriptions because of the complexity of the natural language and limited number of available structured medical reports for training. \citet{textualjustification2019} utilized a justification model on top of the diagnosis network to provide textual and visual reasons for the predictions. In order to alleviate the problem of the duplication of sentences from the training set, a visual word constraint loss was proposed. This approach was validated for breast mass classification and the textual justifications were evaluated using a sentence dataset that described shape and margin of the masses. \citet{Zhang2019PathologistlevelIW} proposed an interpretable framework for diagnostic pathology consisting of three cascading networks. The model generating textual justifications is trained using images and associated diagnostic reports. Five cellular features were outlined in a paragraph by two experienced pathologists. Each paragraph was rephrased by two doctors and two trained students resulting in five paragraphs per image. This method was validation on bladder cancer pathology slides. These methods have an added annotated costs because they require the medical reports to be re-formatted in a structured and concise way.

\subsubsection{Emergent Language}
Symbolic emergent language is the symbolic representation that is learnt spontaneously during communication between two different entities. The generated symbolic expressions can be traced back to the input to provide a layer of interpretability. A symbolic segmentation framework based on emergent language was proposed by \citet{symbolicsegmentation} for interpretable segmentation. It consists of a U-Net like architecture and two Long Short-Term Memory (LSTM) agents, a Sender and a Receiver. The sender receives segmentation mask from the U-Net and generates a symbolic sentence. The receiver works in co-operation with the sender and generates the segmentation mask based on the symbolic sentence. Hence, the network learns a language inorder to communicate between the two modules. This method was validated on brain tumor segmentation in FLAIR images. The generated symbolic expressions are interpretable in terms of tumor characteristics. \citet{Chowdhury2021EmergentSL} used an emergent language based CNN comprising of a sender, symbol generator and a receiver to perform immune cell marker-based classification and chest X-ray classification. The generated symbols correspond to representative information in the input images.

\begin{figure}[h!]
\centerline{\includegraphics[width=1\columnwidth]{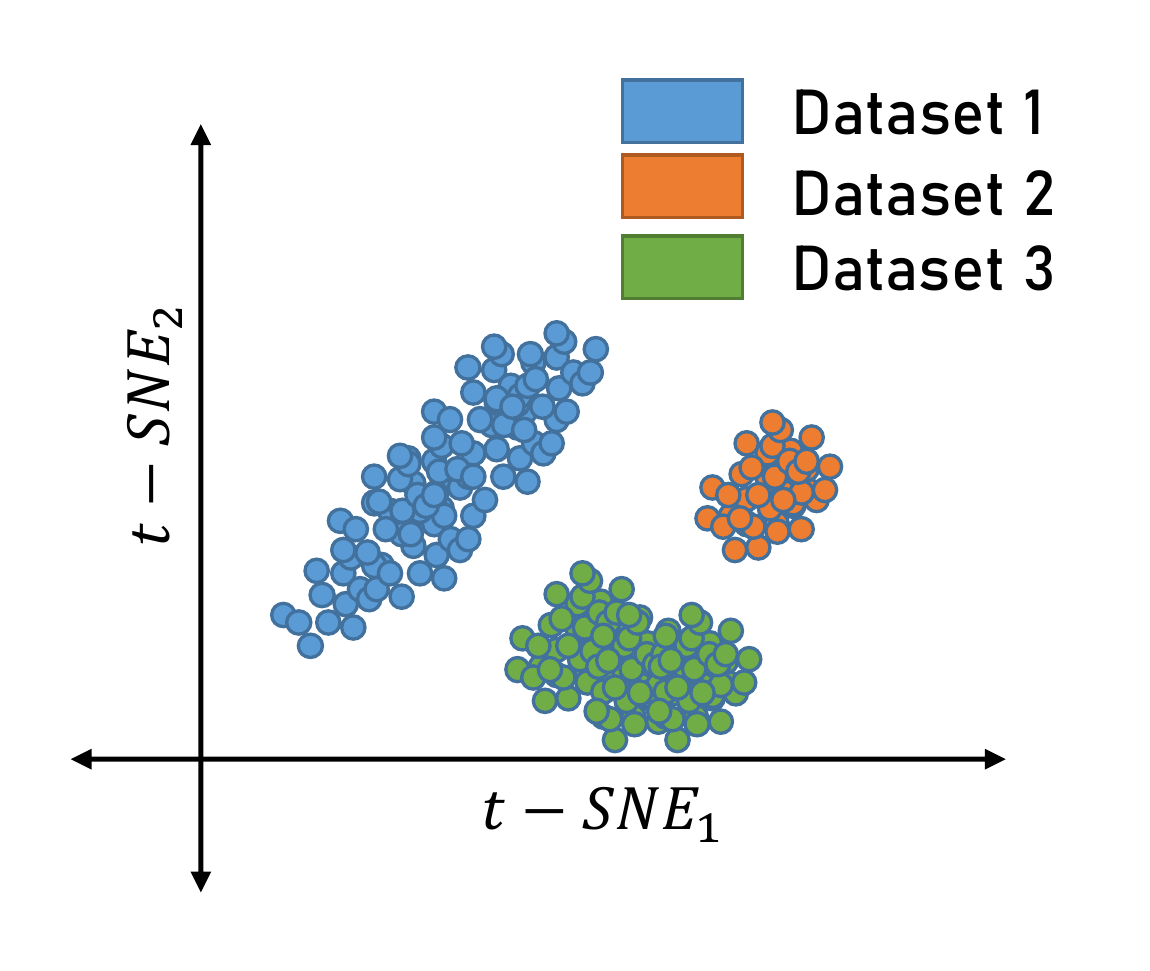}}
\caption{Low dimensional visualization of a fully connected layer of a CNN using t-distributed stochastic neighbor embedding (t-SNE) for interpretability. This plot shows dataset bias in the DL model due to difference in acquisition. }
\label{figure: t-sne}
\end{figure}

\subsection{Latent Space Interpretation}
\label{latentinter}
Latent Space of a convolutional neural network refers to the compressed representation of the input image. Disentanglement representation of the latent space aims to model salient factors of variation in the data independently \cite{higgins2018definition}. An autoencoder is used for non-linear dimensionality reduction and it consists of two parts, an encoder and a decoder. The encoder converts the high dimensional input image into a compressed representation in the bottleneck layer. The latent space in the bottleneck layer is then used to reconstruct the input image by the decoder. An interpretable latent space aims to capture the salient characteristics of the data that can be understood by humans. Variational Autoencoders (VAE) aim to reconstruct the input by learning a continuous latent representation based on variational Bayesian inference \cite{kingma2014autoencoding}. VAE are generative models that also aim to achieve the disentanglement of latent space. 

\citet{biointerpret2020} utilized the latent space of VAE for cardiac resynchronisation therapy response prediction using fully connected layers. Secondary Classifiers are then trained using a part of latent space in order to incorporate clinical domain knowledge and disentangle the latent space. It is ensured that a part of latent space is not used by secondary classifiers so that it can account for unknown contributing factors. \citet{Yang2019DomainAgnosticLW} proposed a framework for cross-modality liver segmentation that generated domain agnostic and anatomy preserving disentangled representation. Visualization of the generated domain agnostic and cross-modality images showed that the disentangled representation helps in the interpretation of the model. \citet{Chen2020} proposed a method called concept whitening that decorrelates and normalizes the latent space and can be used as an alternative to batch normalization for any layer without loss in performance. Pre-defined concepts are then associated with the axes of the disentangled latent space to provide insight into how these concepts are gradually learned. This method was validated for skin lesion diagnosis. \citet{Biffi2020ExplainableAS} visualized the anatomical variability encoded in the latent space to make the classification task interpretable. \citet{CloughTCAV2019} interpolated the latent space of VAE in the direction of concept activation vector inorder to visualize the behaviour of the classifier. However, the quality of the generated images is not high enough to observe the changes. 

The feature space of the CNN consists of hundreds of features even in the bottleneck layers of segmentation networks and fully connected layers of the classification networks. It is not possible to interpret such a high dimensional feature space in order to visualize intra- and inter class similarities, outliers and other data structure information. Therefore, there is a need to visualize these features in a low dimensional space such that the significant structure of the data is preserved. Principal Component Analyis (PCA) is a linear technique that utilizes a global covariance matrix to reduce dimensionality and preserve the distances between input samples. \citet{unbox2021} used UNet with ResNet blocks for brain ventricle segmentation from multi-modal MRI images. The latent space from the encoder is projected into low dimensional space $z_l$ using PCA. Multi-layer preceptron model is trained to predict Dice scores using $z_l$ to estimate the dice scores for the entire input space. 2D projections of $z_l$ showed that input images with certain characteristic features are clustered together. The relationship among features in a non-linear high dimensional space cannot be expressed in a linear form. t-distributed Stochastic Neighbor Embedding (t-SNE) is a non-linear and iterative dimensionality reduction method \cite{JMLR:v9:vandermaaten08a}. t-SNE first calculates the probability of similarity of the features in the high dimensional space and then in the low dimensional space. Kullback-Leibler divergence is used to minimize the difference in probability distributions using gradient descent. Hence, t-SNE can be used to visualize the patterns in low dimensional space. \citet{DINSDALE2021117689} used t-SNE plots for detecting the bias in the dataset from different scanners by visualizing the features from the fully connected layer on a two-dimensional plot. t-SNE plots were used to validate the proposed iterative approach for unlearning the scanner information to produce scanner invariant features. \citet{Faust2018VisualizingHD} utilized t-SNE for generating two dimensional plots from CNN's final hidden layer inorder to the visualize the clusters of similar classes. They used t-SNE plots to provide alternate classification output by k-nearest neighbor approach. Projective Latent Interventions (PLIs) for retraining the classifier by back-propagation based on parametric approximation of t-SNE embedding \cite{latentrepresentation2020}. This method was validated for standard plane classification for fetal ultrasound imaging.

\subsection{Attribution Map}
\label{attrmaps}

 The DL models can be explained by highlighting regions of the input image that are relevant for the prediction. These heatmap-based explanations do not offer any information on how these salient regions contribute to the prediction. 
 
\subsubsection{ Layerwise Relevance Propagation} 
\label{lrp}
Layerwise Relevance Propagation (LRP) is a method based on  pixel-wise decomposition of non-linear classifiers that generates a heatmap by evaluating a relevance score \citep{Bach2015OnPE}. The magnitude of contribution of each neuron in a particular layer is identified in terms of relevance score in a backwards fashion, one layer after the other, starting from the last layer. The relevance score ${R_j}^{(l)}$ of the neuron $j$ in layer $l$ is calculated from the relevance score ${R_k}^{(l+1)}$ of the neuron $k$ in the layer $l+1$ using ${R_j}^{(l)}= \sum_{k} \frac{{a_j}^{(l)} \: \times  \: w_{jk}}{\sum_{j} {a_j}^{(l)} \: \times \: w_{jk}} \: \times \:  {R_k}^{(l+1)}$ where $a_j(l)$ represents the activation of a neuron $j$ and $w_{jk}$ represents the weights between the neurons $j$ and $k$. LRP ensures that the total relevance for all the layers starting from the classification output $f(x)$ to the input layer is the same. Under these constraints, a neuron is highly relevant if it has a high activation and a high contribution for a neuron of the next layer that have a high relevance score.

\citet{Boehle2019LayerWiseRP} demonstrated that LRP can be used to explain Alzheimer's disease classification in 3D CNNs with high inter-patient variability and quantitatively showed that LRP relevance maps correlate with clinical knowledge. \citet{Fabian2019} evaluated LRP visualizations quantitatively for Multiple Sclerosis (MS) diagnosis and concluded that LRP focuses not only on the individual lesions but also on conventional MRI biomarkers in MS. \citet{Thomas2019AnalyzingND} utilized LRP to identify physiologically appropriate brain regions corresponding to the cognitive state decoded from fMRI data using recurrent neural networks. \citet{Yan2017} used LRP to highlight relevant information in high dimensional fMRI data. \citet{ComparisonAttribution2019} carried out a quantitative comparison of four attribution methods for Alzheimer's disease classification and showed that LRP and GBP produce the most coherent explanations.

\subsubsection{Class Activation Maps} 
\label{CAM}
Class Activation Maps (CAMs) localize class specific image regions that the model considers important for classification \cite{Zhou2016LearningDF}. In order to generate CAMs, a global average pooling layer is added after the last convolutional layer. The output of global average pooling layer is then linearly combined to produce class predictions. CAMs for each class are then obtained by taking a weighted sum of the last convolutional layer activations. This interpretability comes at a cost of performance as the proposed network cannot incorporate  fully connected layers. \citet{Camalan2021} used CAMs for interpreting the results of the classification model for oral cancer. Multi-Layer Class Activation Maps (MLCAM) is an extension of CAM that can be incorporated at different CNN layers \cite{MLCAM2018}. ML-CAM was used to identify diagnostic features of human gliomas in Confocal Laser Endomicroscopy images.

\begin{figure*}[h!]
\centerline{\includegraphics[width=\textwidth]{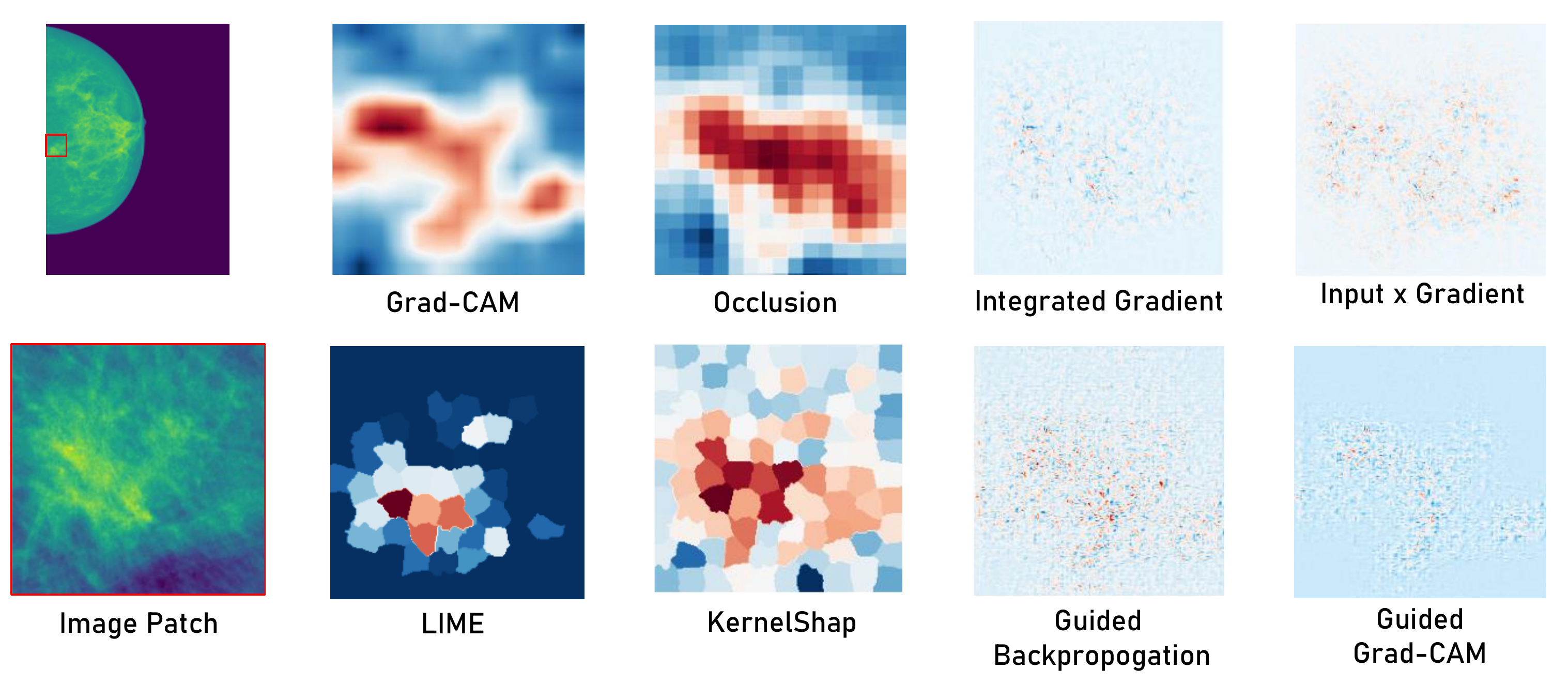}}
\caption{Attribution maps generated by different interpretability methods for explaining a DL model that detects breast mass in an image patch extracted from a digital mammogram. The first column shows the image patch extracted from the mammogram that is provided to the DL model as input. Red color corresponds to important regions and blue color corresponds to irrelevant image regions for classification. }
\end{figure*}

\subsubsection{Grad-CAM} 
\label{GRADCAM}
 Gradient-Class Activation Maps (Grad-CAM) produce visual explanations that do not require re-training or changes to the architecture like Class Activation Maps \cite{Selvaraju2019GradCAMVE}. Grad-CAM allows to explain activations in any layer of the network. The gradient of class c ($y^c$ before softmax) is computed with respect to the feature map activations of a particular layer $A^k$ in order to determine importance $\alpha^{\:c}_{\:k}$ of a particular feature map  for the target class c as $ \alpha^{\:c}_{\:k} = \frac{1}{Z} \: \sum_{i} \sum_{j} \; 
\frac{\partial y^c}{\partial A^k}$. The weighted sum of the activations of the layer $A^k$ with $\alpha^{\:c}_{\:k}$ gives the Grad-CAM output $G^{\:c}$ as: $G^{\:c} = $ ReLU ( $\sum_k \: \alpha^{\:c}_{\:k} \, A^k$). $G^{\:c}$ is then upsampled to match the input image size. 
Grad-CAM's output for last layers in the network lacks fine-grained details. Guided Grad-CAM fuses Guided Backpropagation \cite{Springenberg2015StrivingFS} and Grad-CAM by element-wise multiplication inorder to retreive the fine details. Grad-CAM++ is an extension of Grad-CAM that provides better localization and is capable of detecting multiple object occurrences \cite{Chattopadhay2021}. \citet{gradcamattr2018} used Grad-CAM for the interpretability of DL model for automatic brain tumor grading from MRI data. 

\subsubsection{Integrated Gradient} 
\label{IG}
 Integrated Gradient (IG) method proposed by \citet{IG17} mitigates the saturation problem of gradients. In this method, it is critical to select a baseline  which corresponds to a near-zero score. A complete black image is a suitable choice for a baseline. The gradients are aggregated for all the points occurring in small steps between the input and baseline. IG method satisfies the desirable attribution axioms of sensitivity and implementation variance. \citet{IG17} used attribution maps generated by Integrated Gradient method to interpret DL model for Diabetic Retinopathy (DR) prediction. \citet{IGWargnier} used integrated  gradients for the interpretability of a model for Multiple Sclerosis diagnosis. \citet{SAYRES2019552} conducted a collaborative human-AI study to evaluate the effect of using IG attribution maps for DR grading. Ten ophthalmologists performed the task with and without IG explanations. The accuracy of diagnosis increased for cases with DR when the ophthalmologists used attribution maps.
 

\begin{table*}[H]
\caption{Examples of different methods related to medical image analysis tasks that use attribution maps for the interpretability of DL models.}
\label{tableattribution}
\centering
\begin{tabular}{p{0.35\textwidth} p{0.10\textwidth}p{0.45\textwidth}}
\hline
\Centering{Method} 
& \Centering{Reference} 
& \Centering{Application} 
\\ \hline
Layerwise Relevance Propagation (LRP)
& \Centering{\cite{Boehle2019LayerWiseRP} }
& Alzheimer's disease classification
\\

&\Centering{\cite{Fabian2019} }
& Multiple Sclerosis diagnosis

\\
Class Activation Maps (CAM)
& \Centering{\cite{Camalan2021} }
& Oral cancer classification

\\
Gradient-Class Activation Maps (Grad-CAM)
& \Centering{\cite{gradcamattr2018}} 
& Automatic brain tumor grading\\
& \Centering{\cite{PANWAR2020110190}} & Detection of COVID-19 from Chest X-ray and CT scans

\\
Integrated Gradient (IG) & \Centering{\cite{IG17}} & Diabetic Retinopathy (DR) prediction \\
 & \Centering{\cite{IGWargnier}} & Multiple Sclerosis classification 
 
 \\


Occlusion & \Centering{\cite{KERMANY20181122}} & Diagnosis of age-related macular degeneration and diabetic macular edema in OCT images \\
& \Centering{\cite{natureocclusion2019}} & Diagnosis of Alzheimer's disease

\\

Local Interpretable Model-agnostic \linebreak Explanations (LIME) & \Centering{\cite{ParkinsonLime2020}} & Parkinson's disease detection 
\\
& \Centering{\cite{Seah2019ChestRI}} & Congestive heart failure prediction 

\\

kernel SHAP (Linear LIME + Shapley values) & \Centering{\cite{dermaordeep}} & Skin cancer detection \\
& \Centering{\cite{Zhu2019GuidelineBasedAE}} & Lung nodule classification

\\
Trainable Attention & \Centering{\cite{attentionskin2019}} & Melanoma recognition \\
& \Centering{\cite{attentionYang2020GuidedSA}}& Classification of breast cancer  microscopy images \\
& \Centering{\cite{attentionjetley2018learn}} & Organ segmentation in 3D abdominal CT scans

\\
SmoothGrad & \Centering{\cite{Ghorbani2020}} & Identification of cardiac structure,  estimation of \linebreak cardiac function and  prediction of systemic phenotypes from Echocardiography \\
& \Centering{\cite{SmoothgradERpred}} & Classification of estrogen receptor status from breast MRI

\\
Guided BackPropagation (GBP) & \Centering{\cite{lungGBP2017}} & Lung adenocarcinoma classification 

\\

DeepLIFT (Learning Important FeaTures) & \Centering{\cite{Lopatina2020}} & Diagnosis of Multiple Sclerosis

\\

Deep SHAP (DeepLIFT + Shapley values) & \Centering{\cite{KATZMANN2021141}} & Breast lesion classification, lung lesion classification \\
& \Centering{\cite{Ravi2020ViDiDV}} & COVID and Pneumonia classification from chest X-rays

\\

Deep Taylor Decomposition (Deep Taylor) & \Centering{\cite{imageretreival2020}} & Content-based image retrieval for pleural effusion condition in Chest X-Ray images

\\

Multi-Layer Class Activation Maps (MLCAM)
& \Centering{\cite{MLCAM2018}}
& Feature localization
for Confocal Laser Endomicroscopy Glioma images 
\\

Expected Gradients & \Centering{\cite{DeGrave2021}} & COVID-19 detection 

\\

Contextual Decomposition Explanation \linebreak Penalization (CDEP) 

& \Centering{\cite{Rieger2020InterpretationsAU}} &  Skin Lesion Classification 

\\

\hline

\end{tabular}
\end{table*}

\subsubsection{Perturbation Based Methods}
\label{perturbmethods}
Perturbation based approaches investigate the effect of altering different parts of the input image on the model's predictions. Occlusion method produces attribution map by perturbing the image in a systematic manner to observe the effect on the output. The important parts of the image will have a strong effect on the output when they are altered. Multiple inferences need to be performed for the same input image. It can be computationally expensive to generate occlusion attribution maps if small portions of the image are perturbed. The resolution of the attribution map is constrained by the choice of the patch size that is altered at a time. Occlusion was first utilized by \citet{DeConvolution} for the interpretability of deep neural networks.  \citet{KERMANY20181122} used occlusion maps for interpretability of a model developed for diagnosing age-related macular degeneration and diabetic macular edema in optical coherence tomography images. \citet{natureocclusion2019} utilized occlusion maps for interpretability of a model developed for the diagnosis of Alzheimer's disease. 

Local Interpretable Model-agnostic Explanations (LIME)  can be used to explain a model by approximating its  behaviour locally using a surrogate weighted regression model \cite{Ribeiro2016WhySI}. The training data for LIME is generated by randomly turning a subset of superpixels on or off. Each superpixel corresponds to an input feature and the model's predictions for these perturbed images are the target values. Each perturbed image is weighed by its similarity with the input image using cosine distance metric. A weighted regression model is trained to estimate the feature importance. \citet{ParkinsonLime2020} developed a model for the detection of Parkinson's disease in DaTSCANs and utilized LIME for interpretability. \citet{Seah2019ChestRI} performed a comparative analysis of LIME and their proposed attribution method called Generative Visual Rationale for interpretability of a model predicting congestive heart failure in chest radiograph. 

\citet{LundbergSHAP2017} proposed kernel SHAP (Linear LIME + Shapley values)  that is similar to LIME except that weights  of the regression surrogate model are given by $ \pi_{x^{'}} (z^{'}) = \frac{M-1}{(\frac{M}{\lvert \:z^{'} \rvert}) \cdot (M - \lvert z^{'} \rvert)}$ instead of cosine distance metric. Here, M represents the number of features and $\lvert z^{'} \rvert$ represents the number of non-zero features in the input. \citet{dermaordeep} used kernel SHAP and Grad-CAM to show that model with similar performace use completely different rationales for diagnosing skin cancer. 

Pertubation based methods can alter parts of the image that are not understandable in clinical terms. Therefore, there is a need for meaningful perturbations. \citet{Zhu2019GuidelineBasedAE} proposed Guideline-based Additive eXplanation (GAX) that mitigates this problem by first generating features using rule based segmentation and anatomical irregularities according to the set guidelines. Then, a perturbation based analysis is performed to obtain understandable explanations in terms of feature importance. GAX method was validated for lung nodule classification. \citet{UzunovaVAE2019} generated meaningful perturbations for explaining a classification network for OCT images and brain lesion MRIs by replacing the perturbed area with the healthy equivalent generated using variational autoencoders.

\subsubsection{Trainable Attention}
\label{trainattention}
Trainable attention modules enable the model to focus on important regions of the image and help to suppress the irrelevant background information during training. \citet{attentionjetley2018learn} proposed a method for trainable attention that can be incorporated into standard CNN pipelines. This attention mechanism not only improves classification performance but also produces a fine-grained attribution map that focuses on key parts of the image. The attention module calculates a compatibility score $c^s_i$ between the local features $l^s_i$ obtained from the convolutional layer $s$ where $i$ represents the spatial location within $s$ and the global features $g$ obtained from the final convolutional layer. The attention weights $\alpha_i^s$ can be obtained from the compatibility scores using a softmax function : $\alpha_i^s = \: \frac{exp\:(\:c^s_i\:)}{\sum^n_j \: exp\:(\:c^s_j\:)}$. The final output of the attention module $g^s_a$ is calculated by taking a weighted sum of local features using attention weights as $ g^s_a = \sum^n_{i=1} \: \alpha_i^s \cdot  l^s_i$. The output of the attention modules $g_a$ from various layers are then concatenated and fed into a fully connected layer to perform classification. The attention module tries to ensure that the compatibility score is high only when the image patch contains the part relevant for classification. The greatest benefit is achieved by using these attention modules relatively late in the CNN pipeline. \citet{attentionskin2019} incorporated attention modules in the network to generate attention maps that highlighted image regions relevant for melanoma recognition. The addition of the attention modules resulted in an improvement in performance. \citet{Li2018TellMW} proposed  Guided Attention Inference Network (GAIN) for generating attention maps for weakly supervised tasks by incorporating attention mining loss and a method for improving these attention maps by utilizing extra supervision when available. \citet{attentionYang2020GuidedSA} utilized this guided soft-attention mechanism for the classification of breast cancer microscopy images. 

\citet{attentionSchlemper2019AttentionGN} proposed attention gates that can be incorporated into standard CNN pipelines. These attention gates are more specific to local regions and achieve superior performance compared to global feature vectors utilized in \cite{attentionjetley2018learn}. The proposed attention gates can also be utilized in  segmentation networks unlike in \cite{attentionjetley2018learn}. The use of attention gates demonstrates superior classification performance in scan plane detection for the screening of fetal ultrasound. A gain in performance was achieved by incorporating attention gates in a standard U-Net for the segmentation of multiple organs in 3D CT abdominal dataset. The fine grained attention maps can be visualized to interpret the predictions. \citet{attentionLi2019AttentionDF} used attention gates in Dense-U-Net for breast mass segmentation in mammograms. \citet{attentionSAUNet2020} used spatial and channel-wise attention in the decoder of the proposed U-Net architecture inorder to visualize features learnt at every resolution.

\subsubsection{Other Methods}
\label{othermethods}
 Saliency maps proposed by \citet{Simonyan2014DeepIC} highlight  regions of the input image that the deep neural network considers important for prediction by computing the gradient of the output with respect to the input pixels using backpropagation. Gradient $*$ Input has an advantage that it allows to make use of the input information for better visualization  \cite{shrikumar2017just}. SmoothGrad is a technique that adds noise to the image to generate new images and produces a final attribution map by taking an average of the attribution map of these similar images \cite{smilkov2017smoothgrad} . \citet{Ghorbani2020} used SmoothGrad with simple gradient method for interpretability of DL model applied to echocardiography. Respond-CAM  takes a weighted average of the feature maps with their corresponding gradients and generates attribution map \cite{Zhao2018RespondCAMAD}. Respond-CAM demonstrated better performance when compared with Grad-CAM for 3D biomedical images. \citet{SALEEM2021104410} proposed an extension of method \cite{stergiou2019saliency} for generating attribution maps independent of gradients. This method was evaluated qualitatively and quantitatively for brain tumor segmentation network. 
DeConvNet consists of a series of deconvolution and unpooling layers and produces an attribution map by setting the negative gradients to zero during the backward pass \cite{DeConvolution}. Guided BackPropagation (GBP) is a modification of DeConvolution approach \cite{Springenberg2015StrivingFS}. It works by computing the gradient and setting the negative gradients to zero during backpropagation. \citet{lungGBP2017} used GBP for interpreting decisions made by CNN for lung adenocarcinoma classification. \citet{WICKSTROM2020101619} utilized GBP for interpretability of CNN for colorectal polyps segmentation. \citet{kindermans2018learning} showed that LRP, Guided-BackProp and DeConvNet do not produce theoretically correct explanations for a single-layer architecture i.e. linear models. They proposed PatternNet and PatternAttribution explanation methods that provide correct signal visualization for linear models. 
 
\citet{shrikumar2017just} proposed DeepLIFT (Learning Important FeaTures) explanation method that assigns importance to the neurons based on the activation of each neuron compared to its reference activation. \cite{Simonyan2014DeepIC, DeConvolution, Springenberg2015StrivingFS} can produce misleading results when the local gradient is zero. DeepLIFT does not face this problem because difference from the reference value maybe non-zero even when the local gradient is zero.  \citet{LundbergSHAP2017} proposed Deep SHAP that approximates SHAP values by adapting DeepLIFT. \citet{Montavon_2017} proposed an attribution method called Deep Taylor Decomposition (Deep Taylor) that applies taylor decomposition in a layer-wise manner. \citet{imageretreival2020} utilized the attribution maps generated by Deep Taylor method for an automatic content-based medical image retrival system.

Prior knowledge can be utilized to constrain explanations during training to improve model performance and produce high quality attribution maps \cite{Erion2021}. \citet{Rieger2020InterpretationsAU} introduced Contextual Decomposition Explanation Penalization (CDEP) to incorporate explanation error in the loss function and demonstrated that the Grad-CAM heatmaps with CDEP regularization ignore confounding variables during skin lesion classification. \citet{DeGrave2021} used Expected Gradients method \cite{Erion2021} that utilises attribution prior to generating attribution maps with desirable properties like smoothness and sparsity for the detection of COVID-19 from chest radiographs.

\begin{figure*}[h!]
\centerline{\includegraphics[width=\textwidth]{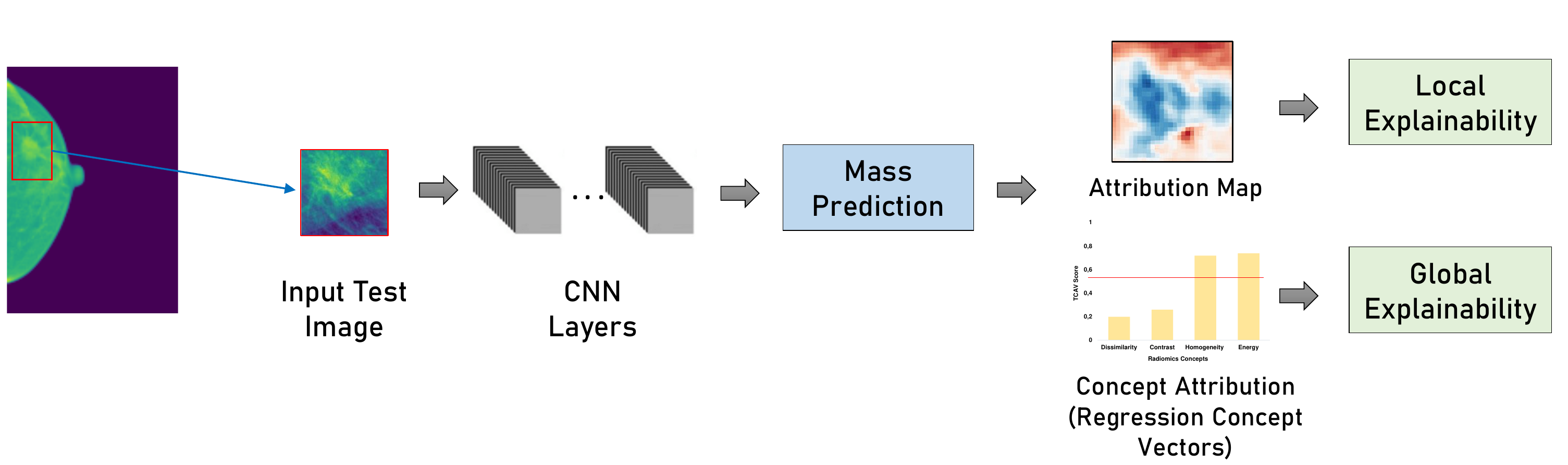}}
\caption{Post-hoc interpretability for mass detection in mammograms. Attribution map provides explanation for each individual test image by highlighting the regions of the image that the model considers important and testing with regression concept vector method provides global interpretability by quantifying the influence of radiomics features on the prediction for each class.}
\label{figure: overview}
\end{figure*}

\subsection{Anatomical Prior}
\label{prior}
Deep Neural Networks for medical image analysis problems can be made interpretable by incorporating task specific structural information/anatomical prior in the workflow. Midline shift (MLS) in brain is an important characteristic feature that can be used for the diagnosis of traumatic brain injury, brain tumors and some other brain abnormalities \cite{Liao2018}. A simple approach for predicting MLS could employ a CNN for direct diagnosis using MRI slices. It would be difficult to interpret the predictions of such a model as several key points need to be identified for MLS prediction and their relative information is also important. \citet{midline2019} proposed an interpretable approach by first estimating the mid-line exploiting the structural knowledge and consequently predicting Mid-Line Shift (MLS) in a brain MRI from the estimated curve. Segmentation models can be trained to localize region of interest so that the network learns only from the specified region to enhance interpretability. The interpretability of the trained segmentation network then also comes into question. CNNs are more inclined to learn texture features and introducing a shape bias can improve robustness \cite{robustshape}. Anatomical shape prior can also enhance interpretability of segmentation networks. \citet{attentionSAUNet2020} introduced Shape Attentive U-Net that includes a secondary shape stream  parallel to the original image stream in the proposed architecture and also generates a shape attention map that can be used for interpretability.

\subsection{Internal Network Representation}
\label{internalnetworkrepresentation}
It can be beneficial to visualize the different features being learnt by the filters in a CNN to understand the features the network is using to arrive to a particular decision. These visualizations are meaningful when dealing with natural images, as different filters can represent peculiar structures and natural appearances. However, the structures and patterns in medical images are not very obvious and hard to interpret. 
Activation Maximization detects the patterns that trigger a particular neuron in a deep network by transforming the pixels of a randomly initialized image to maximize the activation of the neuron keeping the weights of the network and desired output constant \cite{visualization_techreport}. Network Dissection is an interpretability method that quantifies the activation and concept alignment by measuring the intersection over union score between the pixel-wise concept annotations and the corresponding activation of the neuron \cite{Bau2017NetworkDQ}. \citet{natekar2020} explored Network Dissection interpretability method to understand information organization in the DL model for brain tumor segmentation. They also used Activation Maximization with regularization to identify different concepts learnt by the model.  

\begin{table*}[H]
\caption{The table summarizes different methods for the interpretability of DL models for medical image segmentation.}
\label{tablesegmentation}
\centering
\begin{tabular}{p{0.15\textwidth} p{0.10\textwidth} p{0.40\textwidth}p{0.25\textwidth}}
\hline
\Centering{Interpretability Method Category} 
& \Centering{Reference}
& \Centering{Details} 
& \Centering{Application} 
\\ \hline
\Centering{Attribution Map}

& \Centering{\cite{SALEEM2021104410}}

& Visual explanations that are  independent of \linebreak gradients for 3D segmentation networks.   

& Multi-Modal MRI Brain Tumor  Segmentation

\\

\Centering{}

& \Centering{\cite{WICKSTROM2020101619}}

& Guided Backpropagation \cite{Springenberg2015StrivingFS} method for \linebreak identifying important pixels. 

& Colorectal Polyps Segmentation

\\

\Centering{}

& \Centering{\cite{attentionjetley2018learn}}

& Trainable soft attention mechanism that highlights salient parts of the  image. 

& Organ Segmentation in 3D  \linebreak abdominal CT scans

\\
\Centering{Concept Attribution}
& \Centering{\cite{DeepDreams2019}}

& Radiomics features are used to  evaluate robustness and sensitivity of deep segmentation networks.

& Liver Tumor Segmentation from CT scans

\\
\Centering{}

& \Centering{\cite{DTCAV2021}}

& Automated concept discovery \cite{D-TCAV} and superpixels are used to employ TCAV method \cite{Kim2018InterpretabilityBF} for providing global  explanations to interpret different  cardiac conditions. 

& Cardiac MRI Segmentation
\\

\Centering{Language Description}
& \Centering{\cite{symbolicsegmentation}}

& Unsupervised symbolic emergent  language \linebreak explanations that are generated using two LSTM agents. 

&  Brain Tumor Segmentation \linebreak using FLAIR images

\\

\Centering{Internal network Representation}

& \Centering{\cite{natekar2020}}

& Network Dissection \cite{Bau2017NetworkDQ} and  Activation \linebreak Maximization \cite{visualization_techreport} methods for understanding the concepts learned by different filters in the CNN.

& Multi-modal MRI Brain Tumor Segmentation

\\


\Centering{Latent Space Interpretation}

& \Centering{\cite{unbox2021}}

& Low dimensional representation of the latent space for identifying clusters of cases with similar \linebreak performance. 

& Multi-modal MRI Brain  \linebreak Ventricle Segmentation

\\ 

\Centering{}
& \Centering{\cite{Yang2019DomainAgnosticLW}}

& Disentangled latent space  representation for the interpretability of the  segmentation network. 

& Cross Modality Liver \linebreak Segmentation

\\ 

\hline
\end{tabular}
\end{table*}

\section{Evaluation of Interpretability Methods}
As evident in the previous section, several efforts are being made to explore interpretability methods for understanding the inherent black-box nature of the DL algorithms. It is difficult to determine which method works better for understanding the deep neural networks for a particular medical imaging application, as explanations can be subjective. Explanations also depend on a particular application and the context of their utility. The uncertainty of DL models is also translated to their interpretability, and there is no ground truth available for explanations. All these characteristics make the evaluation and quantification of the quality of interpretability methods a great challenge. \citet{doshivelez2017rigorous} proposed three different categories for the evaluation of interpretability methods. Application-grounded evaluations involve experts for a specified application e.g. doctors for diagnosis. Human-grounded evaluations involve lay humans to test general quality of explanations. Finally, functionality-grounded evaluations involve proxy tasks instead of humans for evaluating the quality of explanations. Functionality-grounded evaluation metrics which do not involve human interaction are desirable for interpretability due to time and cost constraints.

\begin{table*}[H]
\caption{Examples of different evaluation strategies used to ensure that the explanations produced by various interpretability methods are trustworthy.}
\centering
\begin{tabular}{p{0.25\textwidth} p{0.10\textwidth}p{0.50\textwidth}}
\hline
\Centering{Evaluation} 
& \Centering{Reference}
& \Centering{Details} 
\\ \hline
\Centering{Evaluation in a Clinical \linebreak Setting}
& \Centering{\cite{Holzinger2020MeasuringTQ}}
& System Causability Scale (SCS) can be used by medical doctors to assess the quality of the interpretability methods.

\\
& \Centering{\cite{SAYRES2019552}} & Ten ophthalmologists performed Diabetic Retinopathy (DR) grading in three scenarios: using AI model with integrated gradient explanations, using only AI model and in an unassisted setting. 

\\
\Centering{Evaluation using Imaging Biomarkers}
&  \Centering{\cite{Fabian2019}}
& LRP attribution maps for Multiple Sclerosis prediction were associated with established biomarkers. 

\\
& \Centering{\cite{Boehle2019LayerWiseRP}}
& LRP attribution maps for Alzheimer's disease prediction correlate with hippocampal volume which is a known biomarker of the disease. 

\\

\Centering{Evaluation against Adversarial Perturbations}
& \Centering{\cite{Ghorbani_Abid_Zou_2019}}
& Input images with small perturbations that are visually indistinguishable produce strikingly different attribution maps using DeepLIFT and Integrated Gradient method. 

\\

\Centering{Quantitative Evaluation of Attribution Maps}
& \Centering{\cite{Adebayo2018}}
& Grad-CAM and Gradient method pass the model parameter randomization test and data randomization test.

\\
& \Centering{\cite{dermaordeep}}
& Reproducibility, model dependence and sensitivity tests for kernel SHAP and Grad-CAM attribution maps for skin cancer classification.

\\
& \Centering{\cite{sanitycheck2020}}
& Retraining the model and applying slight translations to assess attribution maps  for explaining DL model for osteoporotic fracture discrimination in CT scans,

\\
& \Centering{\cite{jin2021map}}
& Modality Importance and Modality Specific Feature Importance metrics were used for the evaluation of 16 methods that generate attribution maps for a brain tumor classification task.

\\
& \Centering{\cite{explainabilityregression2020}}
& Area Over Perturbation Curve (AOPC) metric \cite{AOPC2017} was used for the quantitative evaluation of attributions maps for interpretability of a CNN for fetal head circumference estimation from ultrasound images.

\\
& \Centering{\cite{tjoa2021quantifying}}
& A synthetic dataset comprising of cell images with easily distinguishable features was used to quantify and compare attribution maps.

\\

\Centering{Quantitative Evaluation of Counterfactual Explanation}
& \Centering{\cite{singla2021explaining}}
& Frechet inception distance , counterfactual validity  and foreign object preservation metrics along with two other clinical metrics were used for quantative evaluation of explanations for chest X-ray classification task. 

\\

\Centering{Quantitative Evaluation of Language Description}
& \Centering{\cite{Lee2018AnED}}
& BLEU \cite{Papineni2002BleuAM}, ROUGE-L \cite{Lin2004ROUGEAP} and CIDEr \cite{CIDErVedantamZP15} metrics were used for the evaluation of generated natural language descriptions against ground truth reference sentences provided by experts. 
\\

\hline
\end{tabular}
\end{table*}

\subsection{Evaluation in a Clinical Setting}
Application-grounded evaluation in the presence of a medical doctor is necessary to determine if the end-users are satisfied with the provided explanations. System Causability Scale (SCS) assesses the quality of the interpretability methods which can be utilized in the medical domain \cite{Holzinger2020MeasuringTQ}. It is also important to determine if the provided explanations introduce a bias towards a specific class that can result in performance drop. \citet{SAYRES2019552} studied the impact of DL algorithm and Integrated Gradient explanations on physicians for diagnosing Diabetic Retinopathy (DR) in a computer-assisted setting. When Integrated Gradient explanations were used along with model predictions by ten ophthalmologists, it resulted in an over-diagnosis of the disease in  normal cases without DR. This maybe due to the fact that attribution maps generated by Integrated Gradient method provide evidence for the occurrence of the disease and they may not be useful when there is no disease present. The human-AI collaboration that  utilized explanations achieved overall better accuracy. Therefore, there is a need to evaluate the attribution methods in a clinical setting in order to determine if the explanations are useful for the clinicians and they are not introducing bias towards one class.

\subsection{Evaluation using Imaging Biomarkers}
An imaging biomarker can be described as any quantifiable feature or structure in an image that can be used for the diagnosis and prognosis of a disease \citep{Weaver2017}. DL models automatically detect patterns from the data and learn complex functions that are not understandable to the humans. These functions may learn to utilize established biomarkers from the data along with unknown imaging biomarkers for diagnosis. The explanations of DL models can be validated by using known imaging biomarkers to show that the DL model is using clinically relevant features for diagnosis. This validation can foster trust in the utlity of DL models in clinical practice.
\citet{Fabian2019} utilized Layer-wise Relevance Propagation to develop an association between the relevant features that the model uses and the established biomarkers for Multiple Sclerosis. Imaging biomarkers can be used as concepts in order to quantify their influence on the model's prediction using concept activation vector analysis \cite{CloughTCAV2019}. \citet{Boehle2019LayerWiseRP} used LRP to explain deep neural networks for the diagnosis of Alzheimer's disease. It was quantitatively shown that areas that have high relevance correlate well with hippocampal volume which has been identified as a key biomarker of Alzheimer's disease. A high variability in heatmaps between different disease cases showed that LRP maybe used to identify biomarkers for different stages of Alzheimer's disease. A cascade of two autoencoders was employed by \citet{Waldstein2020} to obtain a global representation of 3D Optical Coherence Tomography (OCT) in just 20 numbers which represented global and local intrinsic features. This low dimensional feature representation was then validated by a comparison with the heatmaps of conventional biomarkers obtained by automated image segmentation algorithms as well as using correlation and machine learning regression.  Imaging biomarkers can also be utilized for both qualitative and quantitative evaluation of interpretability methods.

\subsection{Evaluation against Adversarial Perturbations}
Adversarial examples have demonstrated the vulnerabilities present in the state-of-the-art DL solutions \cite{goodfellow2015explaining}. The security of an automated health care solution for diagnosis, treatment planning and treatment follow-up is critical and should not be compromised by input perturbations \cite{finlayson2019adversarial}. Interpretability methods are also susceptible to similar adversarial attacks. \citet{Ghorbani_Abid_Zou_2019} tested the robustness of DeepLIFT and Integrated Gradient attribution maps to small perturbations of input data. The attribution maps of visually indistinguishable input images with similar labels are radically different. Therefore, there is a need to test the robustness of interpretability methods to input perturbations and noise. 

\subsection{Quantitative Evaluation for Trustworthy Explanations}
Explanations need to be robust, sensitive to model and data, and consistent. Qualitative evaluation alone cannot ensure these properties for some interpretability methods. There is a need to evaluate explanations using quantitative metrics to ensure that the interpretability methods hold these desirable properties. Different metrics and evaluation strategies have been proposed to overcome the lack of explanation ground truth. It is not possible to come up with quantitative metrics that are applicable for all interpretability methods \cite{electronics10050593}. 

\subsubsection{Attribution Maps}
 Attribution maps offer explanation by showing the part of the input image that the model considers important for making the decision. It does not explain how the relevant information is being utilized by the model. The explanation heat map for multiple classes may be same. Correct image area maybe highlighted in the heat map even if the prediction is wrong. This can instill a false sense of confidence in the model which can be detrimental for troubleshooting and model deployment in clinical practice \cite{Rudin2019}. The robustness of the attribution maps produced by different methods cannot be solely evaluated qualitatively. \citet{Adebayo2018} utilized model parameter randomization test and data randomization test in order to quantitatively evaluate if the explanations generated by attribution methods are faithful to the model and the data. Guided BackProp and Guided Grad-CAM fail both the tests while Integrated Gradient and Input $*$ Gradient fail the data randomization tests. Only Gradient and Grad-CAM pass both the tests. \citet{dermaordeep} performed reproducibility, model dependence and sensitivity tests for kernel SHAP and Grad CAM attribution maps for skin cancer classification. DL models with similar accuracy can generate strikingly different attribution maps \cite{sanitycheck2020,dermaordeep}. In addition to the tests proposed in \cite{Adebayo2018, dermaordeep}, \citet{sanitycheck2020} assessed the attribution maps  for explaining DL model for osteoporotic fracture discrimination in CT scans by retraining a given model configuration and applying slight translations. It showed that some of the attribution methods behave as edge detectors. \citet{Eitel2019} trained DL models for Alzheimer's disease classification in identical settings in an effort to test the robustness of the attribution maps produced by LRP, Gradient * Input, GBP and occlusion methods. SmoothGrad is sensitive to sample size used for taking the average and LIME is sensitive to the number of perturbed instances as well as the random seeds for superpixels generation \cite{Bansal2020SAMTS}. \citet{explainabilityregression2020} adapted  Area Over Perturbation Curve (AOPC) metric \cite{AOPC2017} for the quantitative evaluation of attributions maps for interpretability of a CNN for fetal head circumference estimation from ultrasound images. Attribution maps produced by Gradient and DeConvNet methods were found to be insensitive to the model. \citet{tjoa2021quantifying} utilized a synthetic dataset comprising cell images with easily distinguishable features in order to quantify and compare attribution maps for explanation of deep neural networks. A quantitative and qualitative evaluation showed that some methods produce attribution maps that are better at localization while some others at edge detection. It also important to evaluate the performance of attribution maps on multi-modal data. \citet{jin2021map} utilized Modality Importance and Modality Specific Feature Importance metrics for the evaluation of 16 saliency maps along with a doctor user study on a brain tumor classification task. Most of the attribution maps were not able to localize modality-specific key features. 

\subsubsection{Other Methods}
 When using TCAV method, a statistical test for a particular concept by training multiple times ensures the stability of the concept. A two-sided t-test of the TCAV scores should be used to evaluate if the null hypothesis is rejected \cite{Kim2018InterpretabilityBF}. Trustworthy Explainability Acceptance metric quantitatively measures the distance between explanations produced by the AI system and the explanations provided by medical experts \cite{trustworthynewmetric2021}. \citet{singla2021explaining} utilized three metrics for the evaluation of the counterfactual explanations for chest X-ray classification : Frechet Inception Distance (FID), Counterfactual Validity (CV) and Foreign Object Preservation (FOP). FID metric assesses visual quality, CV metric quantifies if the counterfactual is compatible with classifier's prediction and FOP metric checks if the patient specific information is retained. They also used clinical metrics namely cardiothoracic ratio and the score for the detection of normal Costophrenic recess to demonstrate the clinical utility of the explanations. \citet{Lee2018AnED} used BLEU \cite{Papineni2002BleuAM}, ROUGE-L \cite{Lin2004ROUGEAP} and CIDEr \cite{CIDErVedantamZP15} metrics for the evaluation of generated natural language descriptions against ground truth reference sentences provided by experts. 

\begin{figure*}[h!]
\centerline{\includegraphics[width=\textwidth]{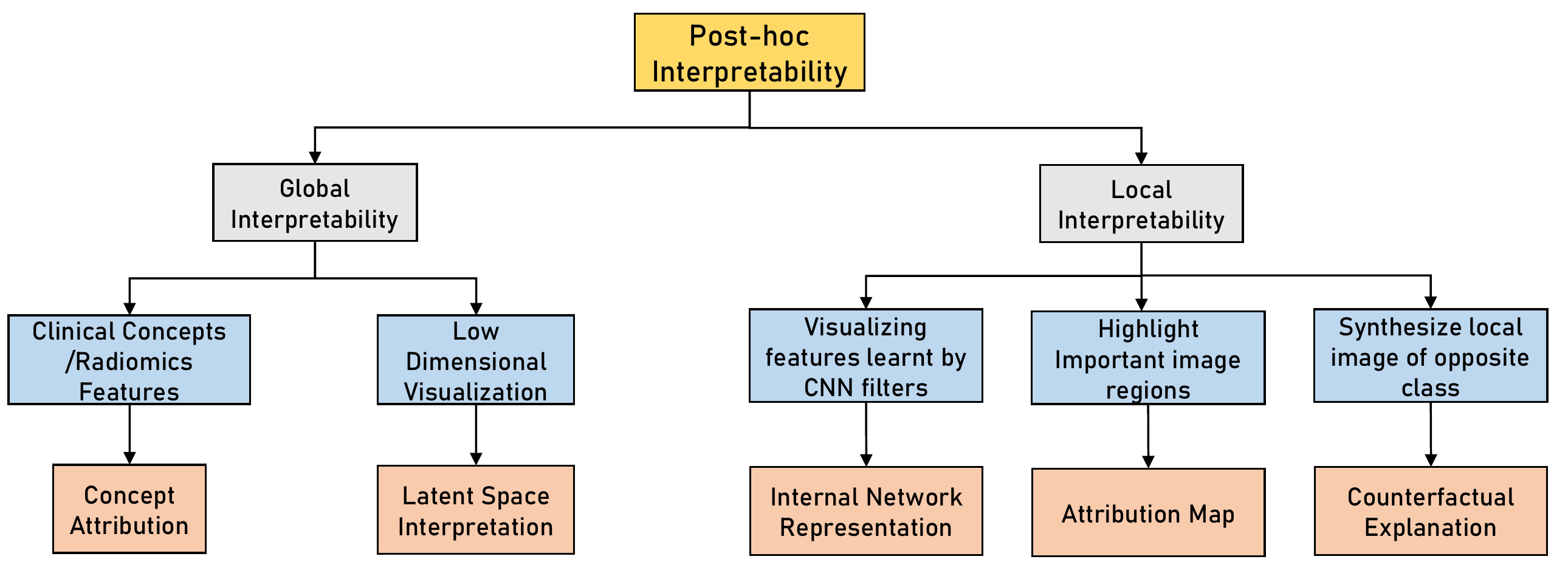}}
\caption{The flowchart shows different methods for post-hoc interpretability for understanding deep learning models for medical image analysis tasks.}
\label{post_hoc_interpretability}
\end{figure*}

\section{Discussion}
Transparency of deep neural networks is an essential clinical, legal, and ethical requirement. We have identified nine different categories of interpretability methods for DL methods. These interpretability methods and their application to medical image analysis problems have been discussed in detail. Evaluation of the interpretability methods is also a critical step in order to validate the utility of the generated explanations and develop insights. Interpretability of DL networks for segmentation is a difficult problem because attribution of a single pixel holds little significance. Hence, there are comparatively few interpretability methods available for segmentation networks. Table \ref{tablesegmentation} shows the various interpretability methods for understanding deep neural networks for medical image segmentation. 

\begin{figure*}[h!]
\centerline{\includegraphics[width=\textwidth]{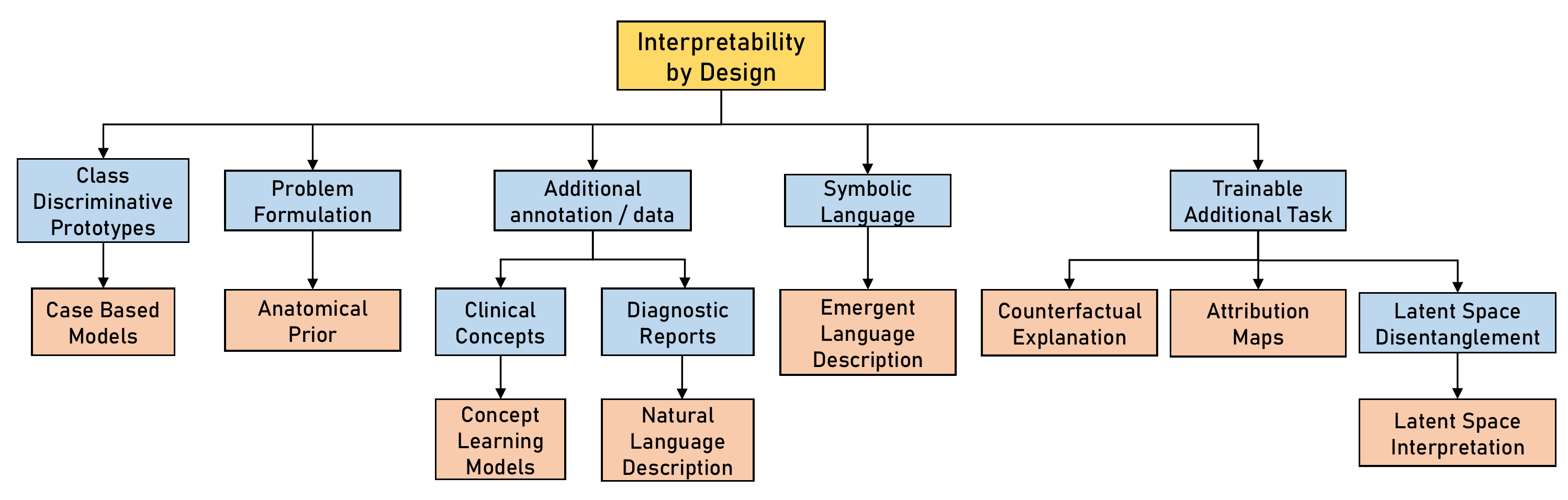}}
\caption{The flowchart shows different methods that incorporate interpretability during the design process of the deep neural network.}
\label{interpretability_by_design}
\end{figure*}

Post-hoc interpretability methods are only approximations and do not accurately depict the correct model behaviour and therefore compromise trust in the explanations \cite{Babic284, Rudin2019}. An additional model for interpretability can lead to an overly complicated decision system because two models may require troubleshooting instead of one. Since post-hoc explanation methods are only approximations, they can not ensure complete accountability \cite{Babic284}. There is a need to carefully evaluate the post-hoc explanations before they can used in the clinical workflow. Figure \ref{post_hoc_interpretability} shows different post-hoc interpretability methods. Interpretability can be introduced during the design process of the deep learning models and can help overcome
the pitfalls of post-hoc interpretability. Figure \ref{interpretability_by_design} shows different methods that incorporate interpretability during the design process. Domain expertise and additional data maybe required to design inherently interpretable models.

\subsection{Performance and Interpretability Trade-off }
It is incorrectly assumed that the gain in performance of DL models comes at a cost of interpretability \cite{unbox2021}. Concept learning models and case based models are interpretable by design and have achieved performance at par with black-box models in medical imaging applications debunking the myth of compromise between performance and interpretability. It depends on researcher's ability to discover the patterns in an interpretable way while at the same time allowing for flexibility to fit the data accurately \cite{Rudin2019}.

\subsection{Discovery of Imaging Biomarkers}
Explainablity methods can hypothesize new imaging biomarkers associated with various diseases.  \citet{Seegerer2020} utilized Layer-wise Relevance Propagation for understanding the intrinsic details of the model. This revealed a number of possible imaging biomarkers for the prediction of Estrogen Receptor status from Haematoxylin-Eosin Images i.e. nuclear and stromal morphology and lymphocyte infiltration. \citet{Zhang2019} introduced a novel architecture for detecting potential imaging biomarkers when only image labels are available. An encoder-decoder architecture was utilized for generating fake normal images to train the discriminator. The discriminator differentiated between normal and abnormal images. At the same time, the difference of encoder-decoder input and output was fed to a CNN classifier. This difference image identified the potential biomarkers. The approach was validated on diabetic retinopathy and skin images with real and stimulated biomarkers respectively. Several efforts have been made to detect imaging biomarkers for Autism spectrum disorder using interpretability methods for DL models \cite{brainbiomarker2018, gametheorybiomarker2019, invertibleASD2019}. In an effort to find an imaging biomarker for Alzheimer's disease from molecular imaging data, \citet{Ding2018} utilized saliency maps for the interpretation of the DL model. The analysis did not reveal any human interpretable imaging biomarker as the model considered the entire brain volume for the diagnosis. This affirms the need for further improving interpretability methods so that we can better understand the DL models in order to detect new imaging biomarkers. These imaging biomarkers after validation can then be used for risk assessment and diagnosis in clinical practice. The skin surrounding a lesion has also been identified as a potential biomarker for skin lesion diagnosis using interpretability methods for DL models \cite{skinbiomarker2019,Tschandl2020HumancomputerCF}. A study on the comparison of diagnostic performance for skin cancer recognition revealed that AI algorithms achieve better accuracy than human experts \cite{Tschandl2019ComparisonOT}. This improvement was particularly prominent for the diagnosis of pigmented actinic keratoses. \citet{Tschandl2020HumancomputerCF} demonstrated that Grad-CAM shows that the CNN pays higher attention to the outside region of the lesion as compared to other categories. This observation is validated by the fact that chronic sun damage is responsible for actinic keratoses and is visible in the surrounding skin region. Medical students were instructed to pay attention to chronic sun damage for diagnosis and this resulted in an overall increase of performance for all categories.

\subsection{Validation of explanations using multi-modal data}
Validation of explanations can be done using genomics and pathology. \citet{Panth2015IsTA} investigated the causal relationship between genetic features and radiomics features by designing an experiment to observe the effect of genetic induction and irradiation. Image features can be correlated with tumor biology using ground truth pathology substrates \cite{SANDULEANU2018349}. \cite{Grossmann2017DefiningTB} studied the association between tumor biology and radiomics features using two independent cohorts with lung cancer. Similar experiments need to be performed for to validate explanations for DL models.

\subsection{Guidelines for using Interpretability methods}
Interpretability of the deep neural networks should be considered as important as the performance. In order to foster trust in the DL solutions for medical image analysis tasks, it is important to involve clinicians in the model design process so that the algorithm becomes interpretable inherently e.g. using multi-modal data (text, images), annotating clinical concepts for diagnosis. Both quantitative and qualitative evaluation of post-hoc interpretability methods should ensure the robustness and faithfulness of the explanations. The insights and explanations provided by the interpretability methods can help discover new imaging biomarkers. Applications-grounded evaluations should be carried out in presence of a clinician to ensure utility of the explanations and eliminate concerns related to bias.

\subsection{Future Directions}
Case-based models and concept learning models are promising interpretability methods that have the potential to be incorporated into clinical workflow in future as they are interpretable by design and they demonstrate performance that is comparable to black-box CNNs. Sanity checks for attribution maps are important to ensure their robustness. Multiple post-hoc interpretability methods can be utilized to understand how the regions highlighted by attribution maps contribute to the final prediction. Multi-modal data consisting of images, texts and genomics data can complement information to increase performance and enhance interpretability. \citet{HOLZINGER202128} proposed that Graph Neural Networks may be utilized to make use of multi-model data and knowledge bases to build a human-in-the-loop interactive explainability solution. These counterfactual graphs can establish causal links between multi-modal feature representation that may comprise of images, text, genomics data etc. using graph structures. There is a need to combine human centered evaluations and quantitative functionality based evaluations to suggest the most suitable interpretability methods \cite{electronics10050593}. 








\section{Conclusion}
The incorporation of deep neural networks in the clinical workflow for medical image analysis tasks is impeded by the vague understanding of the decision-making process. This review paper summaries the technical details, limitations and applications of interpretability methods that are available for achieving some degree of transparency for deep neural networks for medical image analysis tasks. Quantitative and qualitative evaluations are necessary to ensure robust and trustworthy explanations. Application-grounded evaluations in the presence of a medical doctor determine the utility of the interpretability methods and help in detecting if the explanations introduce bias that can result in over-diagnosis or under-diagnosis. Interpretability methods can be validated using imaging biomarkers and they can also aid in discovering new imaging biomarkers.  Case-based models and concept learning models that are inherently interpretable DL models have achieved performance at par with black-box networks. Post-hoc interpretability methods should be utilized with care as they are approximations of the model behaviour and can instill a false sense of confidence in the AI solution.

\section{Acknowledgements}
Authors acknowledge financial support from ERC advanced grant (ERC-ADG-2015 $n^{\circ}$ 694812 - Hypoximmuno), the European Union’s Horizon 2020 research and innovation programme under grant agreement: MSCA-ITN-PREDICT $n^{\circ}$ 766276, CHAIMELEON $n^{\circ}$ 952172, EuCanImage $n^{\circ}$ 952103, TRANSCAN Joint Transnational Call 2016 (JTC2016 CLEARLY $n^{\circ}$ UM 2017-8295), H2020-JTI-IMI2-2020-23-two-stage and IMI-OPTIMA $n^{\circ}$ 10103434.

\bibliographystyle{cas-model2-names}

\bibliography{cas-refs}

\end{document}